\documentclass[english,prl, manuscript]{revtex4}
\usepackage[T1]{fontenc}
\usepackage[latin9]{inputenc}
\usepackage{color}
\usepackage{graphicx}
\usepackage{amssymb}

\makeatletter

\@ifundefined{textcolor}{}
{%
 \definecolor{BLACK}{gray}{0}
 \definecolor{WHITE}{gray}{1}
 \definecolor{RED}{rgb}{1,0,0}
 \definecolor{GREEN}{rgb}{0,1,0}
 \definecolor{BLUE}{rgb}{0,0,1}
 \definecolor{CYAN}{cmyk}{1,0,0,0}
 \definecolor{MAGENTA}{cmyk}{0,1,0,0}
 \definecolor{YELLOW}{cmyk}{0,0,1,0}
 }

\makeatother

\usepackage{babel}

\begin{document}

\title{}

\title{Dynamics of Discontinuous Transitions in Competitive Percolation}

\title{Impact of Single Links in Discontinuous Percolation}

\title{Impact of Single Links in Competitive Percolation}

\author{Jan Nagler$^{1,2}$, Anna Levina$^{1,3}$ and Marc Timme$^{1,2,3}$}

\affiliation{$^{1}$Max Planck Institute for Dynamics and Self-Organization (MPI
DS) Göttingen }

\affiliation{$^{2}$Institute for Nonlinear Dynamics, Faculty of Physics, University
of Göttingen}

\affiliation{$^{3}$Bernstein Center for Computational Neuroscience (BCCN) Göttingen,
Bunsenstr. 10, 37073 Göttingen, Germany.}
\begin{abstract}
\textbf{How a complex network is connected crucially impacts its dynamics
and function \cite{Strogatz2001,Newman2002,Song2005,Song2006}. Percolation,
the transition to extensive connectedness upon gradual addition of
links, was long believed to be continuous \cite{benAvrahamHavlin2001}}
\textbf{but recent numerical evidence on \quotedblbase{}explosive
percolation\textquotedblleft{} \cite{Achlioptas2009} suggests that
it might as well be discontinuous if links compete for addition. Here
we analyze the microscopic mechanisms underlying discontinuous percolation
processes and reveal a strong impact of single link additions. We
show that in generic competitive percolation processes, including
those displaying explosive percolation, single links do not induce
a discontinuous gap in the largest cluster size in the thermodynamic
limit.  Nevertheless, our results highlight  that for large finite
systems single links may still induce observable gaps because gap
sizes scale weakly algebraically with system size. Several essentially
macroscopic clusters coexist immediately before the transition, thus
announcing discontinuous percolation. These results explain how single
links may drastically change macroscopic connectivity in networks
where links add competitively.}\textbf{\textcolor{green}{}}
\end{abstract}
\maketitle
\global\long\def\DCmax{\Delta C_{\mathsf{max}}}
\global\long\def\numax{\nu_{\mathsf{max}}}
\global\long\def\Pgr{p_{\mathsf{gr}}}
\global\long\def\rmi{r^{-}}
\global\long\def\rp{r^{+}}

Percolation, the transition to large-scale connectedness of networks
upon gradual addition of links, occurs during growth and evolutionary
processes in a large variety of natural, technological, and social
systems \cite{Strogatz2001}. Percolation arises in atomic and molecular
solids in physics as well as in social, biological and artificial
networks \cite{Solomon2000,Newman2002,Goldstone2005,Dorogovtsev2008,RozenfeldProteinHomologyExplosive09}.
In the more complex of these systems, adding links often is a competitive
process. For instance, a human host carrying a virus may travel at
any given time to one but not to another geographic location and therefore
infect other people only at one of the places \cite{HufnagelPNAS04,dSouzaNatPhys2009}.
Across all percolating systems, once the number of added links exceeds
a certain critical value, extensively large connected components (clusters)
emerge that dominate the system.

Given the breadth of experimental, numerical, and empirical studies,
as well as several theoretical results and analytic arguments \cite{Bunde1996,BollobasBook,StaufferBook,GrimmettBook},
percolation was commonly believed to exhibit a continuous transition
where the relative size of the largest cluster increases continuously
from zero in the thermodynamic limit once the number of links crosses
a certain threshold. So recent work by Achlioptas, D'Souza and Spencer
\cite{Achlioptas2009} came as a surprise because it suggested a new
class of random percolating systems that exhibit {}``explosive percolation''
\cite{Bohmann2009}. Close to some threshold value, the system they
considered displays a steep increase of the largest cluster size with
increasing the number of links; moreover, numerical scaling analysis
of finite size systems suggests a discontinuous percolation transition.
This study initiated several follow-up works (e.g.~ \cite{Cho2009,RadicciExplosiveScaleFreePRL09,RozenfeldProteinHomologyExplosive09,ZiffExplosive2DLatticePRL09,Cho2010,Friedman2009,Moreira2010,Radicchi2010,Souza2010})
confirming the original results for a number of system modifications.
These in particular support that competition in the addition of links
is crucial; the key mechanisms underlying discontinuous percolation,
however, are still not well understood and the impact of individual
link additions is unknown.

\begin{figure}
\begin{centering}
\includegraphics[width=14cm]{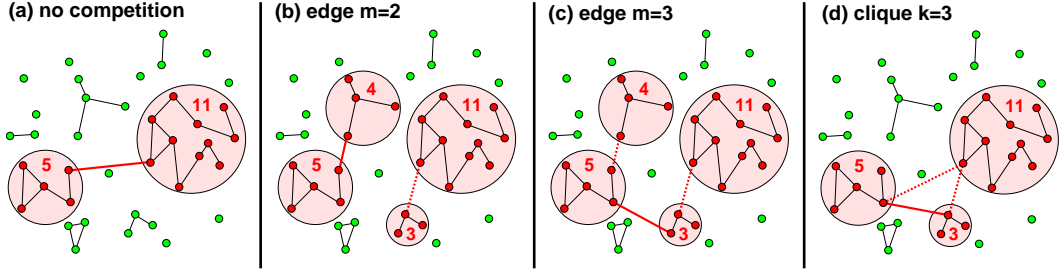} 
\par\end{centering}

\caption{\label{Fig:ModelDescription} Competitive vs. non-competitive percolation
processes. (a) Non-competitive Erdös-Rényi percolation: new randomly
chosen links just add. (b) Edge competition: $m=2$ links compete
with each other and clusters of sizes 4 and 5 win the competition
and join to form a new cluster of size 9 (c) $m=3$ links compete
with each other. Clusters of sizes 3 and 5 join. (d) Clique competition
($k=3$): three links within a clique compete. Clusters of sizes 3
and 5 join. Throughout all panels, small disks indicate nodes, solid
black lines existing links; large shaded disks indicate clusters entering
the competition with numbers denoting their sizes; red-dashed lines
indicate potentially new, competing links; solid red lines indicate
actual link added. \label{fig:ClassOfCompetitionModels}}

\end{figure}

Gaining one or a few links may have drastic consequences for a network's
growth and its overall dynamics, depending on whether or not such
individual links qualitatively alter the global connectivity of a
network. For instance, spontaneous activity in developing neural circuits
may become persistent after establishing some additional synaptic
connections \cite{BreskinPRL06,SorianoPNAS08}. Similarly, during
beginning pandemics the specific travel patterns of a single infected
person may substantially change the number of infecteds on a time
scale of months \cite{HufnagelPNAS04}.

Here we identify how microscopic single-link additions impact competitive
processes. We find that in generic percolation processes, single links
do not induce macroscopic gaps in the largest cluster size as the
system size $N\rightarrow\infty$. Nevertheless, the gap sizes decay
weakly algebraically as $N^{-\beta}$ with often small $\beta$ such
that gaps are essentially macroscopic, i.e. substantially large even
for systems of macroscopic size $N\approx10^{23}$. Such gaps, induced
by single links, occur at the point of percolation transitions, are
a key signature of discontinuous percolation, and are announced by
several coexisting, essentially macroscopic clusters.

\section*{The nature of discontinuities in competitive percolation processes }

Consider a family of competitive percolation processes where potentially
new links compete with others for addition (Fig.~\ref{Fig:ModelDescription}).
Starting with an empty graph of a large number $N$ of isolated nodes
(no links), links sequentially add in competition with others. For
\emph{edge competition}, for each single-link addition, $m$ potential
links are randomly selected. The link for which the sum of the cluster
sizes containing their two end-nodes is smallest wins the competition
and adds. Intra-cluster links are possible; these can only broaden
the transition compared to disallowing them. For $m=1$, this process
is non-competitive and identical to random Erdös-Rényi percolation
\cite{BollobasBook}, whereas for $m=2$ it specializes to the process
introduced before \cite{Achlioptas2009}. For all $m\geq2$, this
kind of competition promotes that during gradual addition of links
smaller clusters tend to be connected (to form larger ones) before
larger clusters grow. With increasing $m$, the competition becomes
more strongly competitive, because more potentially new links actually
compete.  If $m$ is maximal, all potential links in the network
compete for addition and we have \emph{global competition.}

 Taking the sum of cluster sizes in edge competitive processes appears
somewhat arbitrary because, e.g., taking the product \cite{Beveridge2007},
or, for that matter, any convex function of the two cluster sizes,
has similar competitive effects, cf. \cite{Achlioptas2009}. We thus
consider also \emph{clique competition} that does not suffer from
this ambiguity. For clique competition, randomly draw a fixed number
$k$ of nodes and connect those two of them contained in the two smallest
clusters. Here $k=2$ describes non-competitive random percolation
and for all $k\geq3$ competition has the same principal effect on
changes in cluster sizes as edge competition. We remark that for maximal
possible $k$ we again have global competition.

\begin{figure}
\begin{centering}
\includegraphics[width=9cm]{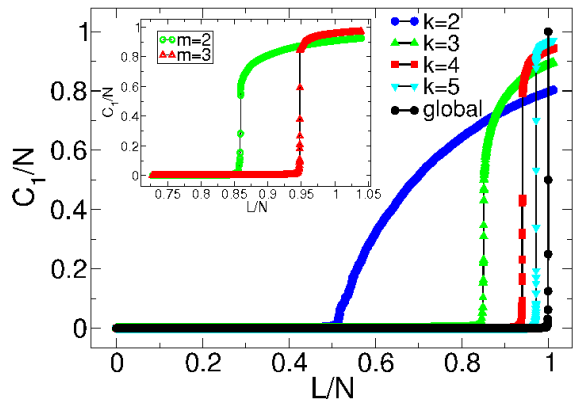} 
\par\end{centering}

\caption{\label{Fig:originalDynamics} Growth of the largest cluster size $C_{1}$
as a function of the number of added links $L$ for non-competitive
($k=2$), competitive, and globally competitive percolation processes
for both edge (inset) and clique competition (main panel); $N=2^{16}$
nodes, quantities on both axes rescaled by system size $N$. Single
realization is displayed for each percolation process.}

\end{figure}

For large finite systems, single realizations of genuinely competitive
processes ($m\geq2$ or $k\geq3$, cf.~Fig.~\ref{Fig:originalDynamics})
exhibit macroscopic $\mathcal{O}(N)$ changes in the size $C_{1}$
of the largest cluster $\mathbf{C}_{1}$. In fact, numerical scaling
studies (Supplementary Fig.~1\textbf{}) confirm that the transition
regime in the plane spanned by $\ell=L/N$ and $c_{1}=C_{1}/N$ shows
an $\mathcal{O}(1)$ change of $c_{1}$ in a region of width $\Delta\ell$
that scales as $N^{-\gamma}$, $\gamma>0$, for large $N$ (cf. also
\cite{Achlioptas2009}). These results may suggest that in the limit
of infinite systems there is a discontinuous $\mathcal{O}(1)$ gap
in the curve characterizing competitive percolation in the $\ell-c_{1}$
plane.

Further investigating the microscopic dynamics of the transition,
however, seeds doubt about any such gap. If the largest gap $\DCmax:=\max_{L}(C_{1}(L+1)-C_{1}(L))$
is macroscopic (extensive), \begin{equation}
\lim_{N\rightarrow\infty}\frac{\DCmax}{N}>0\,,\label{eq:JumpDiscontinuous}\end{equation}
we call such transitions \emph{strongly discontinuous}, otherwise
\emph{weakly discontinuous} (see Supplementary Information\textbf{
}for an exact definition)\emph{.} For weakly discontinuous transitions,
the curve in the $\ell-c_{1}$ plane does in fact not exhibit any
macroscopic gap in the thermodynamic limit. 

\begin{figure}
\begin{centering}
\includegraphics[width=9cm]{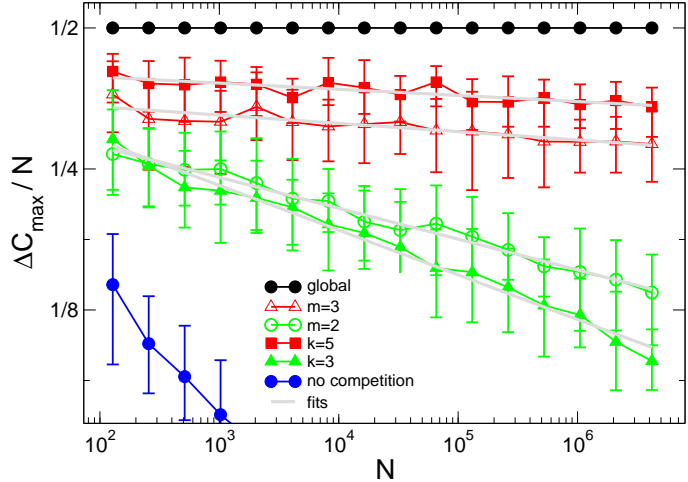} 
\par\end{centering}

\caption{\label{Fig:dcmaxDecay} Gap sizes $\DCmax$ decay algebraically with
system size $N$ for weakly discontinuous transitions. Green symbols:
weakest competition; red symbols: stronger competition. black symbols:\textbf{
}global competition; blue symbols: no competition ($k=2$, $m=1$).
The symbols indicate average values for 50 realizations; error bars
indicate 25\%-quantiles and reflect system-intrinsic fluctuations.
 Solid grey lines are best least-square fits (slopes, $\beta=0.013$
($k=5$) $\beta=0.021$ ($m=3$), $\beta=0.065$ ($m=2$) and $\beta=0.095$
($k=3)$). Black line shows the analytical curve for global competition,
where $\beta=0$. }

\end{figure}

Evaluating the largest jump size $\DCmax$ from extensive numerical
simulations of systems up to size $2^{26}\approx6.7\times10^{7}$
already suggests (Fig.~\ref{Fig:dcmaxDecay}) that it scales algebraically
as \begin{equation}
\frac{\DCmax}{N}\sim N^{-\beta}\label{eq:JumpScalingNbeta}\end{equation}
independent of whether the process is non-competitive, minimally competitive
($k=3$, $m=2$) or exhibits even stronger forms of competition ($k\ge4$,
$m\ge3$). As we find that $\beta>0$ for all such processes, we have
$\lim_{N\rightarrow\infty}\DCmax/N=0$ and thus the transitions are
all weakly discontinuous. The only exception seems to be global competition
where we find $\beta$ indistinguishable from zero and $\DCmax/N\approx0.5>0$
for all $N$ (Fig.~\ref{eq:JumpDiscontinuous}), suggesting a strongly
discontinuous transition. The set of all numerical analyses therefore
suggests that competitive percolation transitions are generically
weakly discontinuous, and single links do not induce a gap in $c_{1}$
in the thermodynamic limit $N\rightarrow\infty$. Nevertheless, as
the gap sizes scale weakly algebraically with system size (\ref{eq:JumpScalingNbeta})
with often small $\beta$ such gaps may still be essentially macroscopic,
i.e. substantially large even for macroscopic systems of large finite
size $N$.

\section*{The impact of single links}

So how can single links actually impact the dynamics of the transition?
For the extreme case of global competition, exact analytical arguments
reveal the occurrence of macroscopic jumps and gives key insights
about the nature of transitions in competitive percolation processes,
that similarly hold for weakly discontinuous transitions (see below):
We label all existing clusters by $\mathbf{C_{i}}$ and their sizes
by $C_{i}=|\mathbf{C_{i}}|$ where the index $i$ enumerates their
size rank such that $C_{1}\geq C_{2}\geq\ldots\geq C_{\numax}$ where
$\numax\leq N$ denotes the total number of existing clusters. For
global competition each newly added link joins the two smallest clusters
in the entire system such that $\mathbf{C}_{\nu_{\mathsf{max}}}+\mathbf{C}_{\nu_{\mathsf{max}}-1}\rightarrow\mathbf{C}'$.
For simplicity of presentation, we choose the system size $N$ to
be a power of 2. This ensures that up to $L_{1}=N/2$ new links only
connect 1-clusters (isolated nodes) to result in new 2-clusters (two
nodes with a single connecting link) such that the maximum cluster
size stays $C_{1}=2$ for all $L\leq L_{1}$. The subsequent $N/4$
links each connect 2-clusters to 4-clusters, keeping $C_{1}=4$ until
$L_{2}=3N/4$. In general, new links added between $L_{n-1}$ and
$L_{n}$ connect $n$-clusters to $2n$-clusters keeping $C_{1}=2^{n}$
where $L_{n}=\frac{(2^{n}-1)}{2^{n}}N$, for all $n\leq\log_{2}(N)$.
 In the final step, at $L=N-1$, the remaining two $\frac{N}{2}$-clusters
join and induce the largest gap \begin{equation}
\frac{\DCmax}{N}=\frac{1}{2},\label{eq:DCmaxGlobal}\end{equation}
analytically confirming the numerical findings (Figs.~\ref{Fig:originalDynamics}~and~\ref{Fig:dcmaxDecay}).
As a consequence, global competition (involving information about
the entire system's state for local link addition) implies a genuine
gap of size $1/2$ in the main order parameter $c_{1}$.

\paragraph{}

For weaker forms of competition, closely related link adding mechanisms
control the cluster joining dynamics. Inspecting the impact of single
link additions onto cluster joining dynamics  in more detail we
identify three distinct mechanisms that may contribute towards increasing
the size $C_{1}$ of the current largest cluster in more general competitive
processes:
\begin{description}
\item [{(i)}] Largest cluster \emph{growth}: the largest cluster itself
connects to a smaller cluster of size $C_{i}<C_{1}$ and grows, $\mathbf{C_{1}+C_{i}}\rightarrow\mathbf{C_{1}}$,
to stay the largest cluster. 
\item [{(ii)}] \emph{overtaking}: two smaller clusters of size $C_{i}\,,C_{j}<C_{1}$
join into one that is larger than the current largest cluster, $\mathbf{C_{i}}+\mathbf{C_{j}}\rightarrow\mathbf{C_{1}}$,
and the originally largest cluster becomes second largest, $\mathbf{C_{1}}\rightarrow\mathbf{C_{2}}$;
\item [{(iii)}] \emph{doubling}: if there are several clusters of maximal
size $C_{1}=C_{2}=..=C_{\nu}$ for some $\nu\ge2$, two of these join,
$\mathbf{C_{i}}+\mathbf{C_{j}}\rightarrow\mathbf{C_{1}}\;\mbox{for some }i,j\in\{1,\ldots,\nu\}$,
creating a new largest cluster of twice the size of the original one. 
\end{description}
For each single link addition, we denote the probability for normal
cluster growth (i)  by $\Pgr$. We say that $\Pgr=0$ if the probability
of normal cluster growth (i) is zero up to the point where only two
clusters are left in the system and normal growth is the only remaining
way the largest cluster could grow at all (see Supplementary Information
for a more formal definition).

As we show in the following, an arbitrary percolation process with\textbf{
$\Pgr=0$ }necessarily exhibits a genuine gap and thus a strongly
discontinuous transition, i.e. $\DCmax/N$ stays positive in the limit
of infinitely large system sizes $N$. As growth (i) is prohibited,
the largest cluster size changes either by overtaking (ii) or by
doubling (iii). During any such percolation process adding a link
never more than doubles $C_{1}$. As a consequence, there is a certain
$L'$ such that $C_{1}(L')$ is larger than $N/3$ but not larger
than $2N/3$. When $\mathbf{C_{1}}$ will be overtaken (or doubles)
one more time at some $L=L'+\Delta L$, the cluster previously largest
becomes the second largest, $\mathbf{C_{1}}\rightarrow\mathbf{C_{2}}$
(or disappears in case of doubling). Thus it is guaranteed that during
percolation two clusters of sizes $C_{1}\ge N/3$ and $C_{2}\ge N/3$
are generated which necessarily join at some time $L>L'$. Therefore,
in any such competitive process, prohibited growth $\Pgr=0$ implies
that the largest gap is macroscopic, \begin{equation}
\frac{\DCmax}{N}\geq\frac{1}{3}.\label{eq:gapLargerOneThird}\end{equation}
Hence, all competitive percolation processes with $\Pgr=0$ display
strongly discontinuous transitions with a strong impact of single
link additions. As we show in the Supplementary Information, such
a gap necessarily occurs at or beyond $\ell_{c}=1$; thus for extremal
competition with $\Pgr=0$ the percolation point, where the largest
cluster becomes macroscopic, does not necessarily coincide with the
point where the largest gap occurs.

\section*{Single links induce gaps in large finite systems}

Nevertheless, many weakly discontinuous transitions still exhibit
essentially macroscopic gaps for large finite systems: We conjecture
that competitive percolation processes in nature (or engineering or
the social world), in particular spatially extended systems with limited
range interactions shall naturally allow the largest cluster to grow,
$\Pgr>0$ (as do all competitive percolation processes for non-global
clique and edge competition) and they generically exhibit weakly discontinuous
(if not continuous) percolation transitions \cite{Binder1987,Goldenfeld1992}.
In specific limiting models analytic mean field considerations yield

\begin{equation}
\frac{\DCmax}{N}\sim N^{-\beta},\;\beta>0,\label{1st}\end{equation}
thus confirming (\ref{eq:JumpScalingNbeta}). For instance, in a model
variant where largest cluster joins with the smallest available with
probability $p$, and otherwise the two smallest clusters join with
probability $1-p$ we analytically find that (see Supplementary Information
for a detailed derivation) \begin{equation}
\beta=1+\frac{\log(2)}{\log[(1-p)/(2-p)]}\approx\frac{p}{2\log(2)}\label{eq:beta_analytic}\end{equation}
for $0\leq p\ll1$ scales roughly linearly with $p$.

\textbf{}%
\begin{figure}
\begin{centering}
\textbf{\includegraphics[width=15cm]{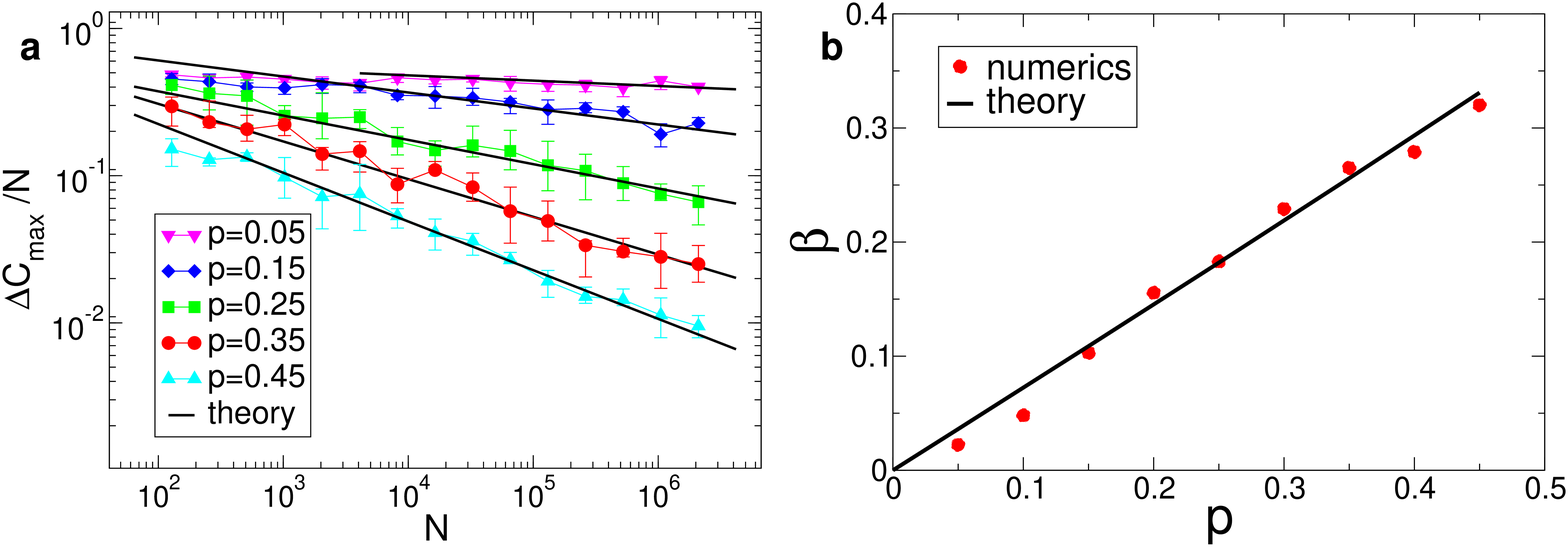} }
\par\end{centering}

\caption{\textbf{} Weakly discontinuous transition in stochastic mixture of
largest cluster growth (with probability $p$) and suppressed growth.
(a) Double-logarithmic plot of $\DCmax/N$ vs. $N$ for different
$p$. The slopes of the theoretical mean field prediction (black lines;
ranging from $\beta=0.036$ ($p=0.05$) to $\beta=0.33$ ($p=0.45$))
asymptotically well fit the gap sizes obtained by numerical experiments
(symbols) . (b) Indeed, the theoretically derived exponent $\beta$
(\ref{eq:beta_analytic}) as a function of $p$ (no fit parameter)
systematically well predicts those found from fitting the data in
(a) (red dots). \label{fig:DCmaxAnalytics}}

\end{figure}

Notably, if largest cluster growth does not occur, $p=0$, we have
$\beta=0$ and  $\DCmax/N>0$ in the thermodynamic limit, consistent
with Eq.~(\ref{eq:DCmaxGlobal}). More importantly, these results
show that even if the largest cluster may grow the slightest, i.e.
for the smallest possible size increase with arbitrarily small $p>0$
the percolation transition is weakly discontinuous, because $\beta>0$
as soon as $p\neq0$. Direct numerical simulations well agree with
our analytical prediction (\ref{eq:beta_analytic}), see Figure \ref{fig:DCmaxAnalytics}.
The finding that $\beta>0$ as soon as $p>0$ is consistent with the
above general result that for arbitrarily small probability $\Pgr>0$
of cluster growth, the percolation transition is already weakly discontinuous,
often with small positive exponents $\beta$ and thus essentially
macroscopic gaps in large finite systems (see numerical example below).\textbf{
}More generally, the results above suggest that any process with non-maximal
competition (including non-maximal edge competition ($m=2$) displaying
{}``explosive percolation'' \cite{Achlioptas2009,Ziff2009,Radicchi2009,Cho2009})
generically displays weakly discontinuous transitions.

\section*{Finite size scaling and coexisting large clusters}

\begin{figure}
\begin{centering}
\includegraphics[width=15cm]{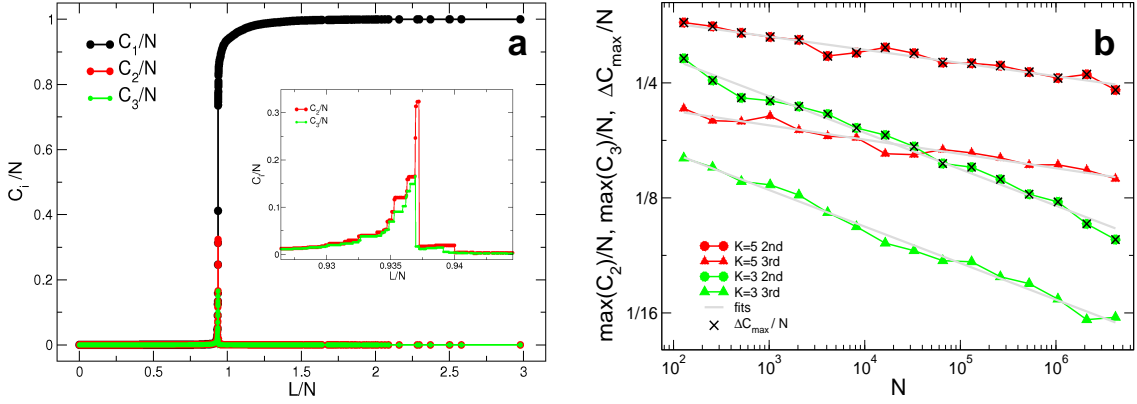} 
\par\end{centering}

\caption{\textbf{ }Several large clusters coexist at a discontinuous percolation
transition. (a) Main panel: Simultaneous emergence of the largest
cluster of size $C_{1}$, the second largest cluster $C_{2}$, and
the third largest cluster $C_{3}$ in a competitive percolation process
(clique percolation, $k=4$, $N=2^{15}$). Inset: blow-up of $C_{2}$
and $C_{3}$ in the region around the transition point. (b) The maximal
sizes of second and third largest clusters as a function of network
size $N$ indicate that they have the same order of magnitude and
the same scaling that is moreover identical to that of $\DCmax$.
In particular, $\max_{L}C_{i}(L)\sim\DCmax\sim N^{-\beta}$ for $i\in\{2,3\}$
with $\beta=0.095\pm0.001$ for $k=3$ and $\beta=0.036\pm0.001$
for $k=5$. The maximum gap size $\DCmax$ ($\times$) in fact exactly
equals the maximum size of the second largest cluster. Thus, there
is no unique large cluster right at the transition even for very large
finite systems. \textbf{\label{fig:nonUniqueCmax}}}

\end{figure}

Further extensive numerical scaling analysis reveals that the gaps
in the generic competitive percolation processes we consider indeed
occur coincident with the point where the largest cluster size is
discontinuous (Supplementary Figure 2). Moreover, immediately before
the transition, not only the largest gap size, but also the second
largest cluster, the third largest cluster etc. appear essentially
macroscopic (Fig.~\ref{fig:nonUniqueCmax}). In particular, the maxmium
second largest cluster generically exactly equals the maximum gap
size, $\DCmax=\max_{L}C_{2}(L)$; see Supplemenary Information for
a derivation. Thus for small $\beta$ the largest cluster is essentially
non-unique, in contrast to standard continuous percolation transitions.
Finally, analytical arguments also demonstrate that the percolation
strength \cite{RadicciExplosiveScaleFreePRL09,Radicchi2010},\textbf{
}defined as the difference in largest cluster size immediately after
and immediately before the gap, exactly equals the size of the second
largest cluster before the transition, which in turn scales with the
same exponent $\beta$ as the gap size (\ref{eq:JumpScalingNbeta}).
Taken together, single link additions induce several new distinctive
features of discontinuous percolation transitions and thus serve as
a key mechanism controlling competitive percolation processes. 

Interestingly, the so-called $k$-cores of the evolving graph, serving
as the key example of the drastic impact of single links in traditional
percolation theory \cite{Spencer2008,Spencer2010}, exhibit dynamics
very similar to that for Erdös-Rényi percolation, even for extreme
processes with $\Pgr=0$. The $k$-core of a graph is the largest
subgraph with minimum degree at least \textbf{$k$}. As numerical
simulations indicate (Supplementary Figure 6), the size of the $2$-core
increases continuously from zero whereas $k$-cores for all $k\geq3$
exhibit a discontinuous jump induced by single link additions. These
results hold for\textbf{ }both Erdös-Rényi as well as competitive
percolation processes. Even for extreme processes with\textbf{ }$\Pgr=0$
the $2$-core is still continuous, but with the location of the transition
moved to larger values compared to the point of percolation. The dynamics
of $k$-cores is thus very similar for competitive and standard,
non-competitive percolation processes, in stark contrast to the dynamics
of the largest cluster size. This is true even though, as shown above,
the latter is also strongly influenced by single link additions.

\section*{Discussion}

These results explain how microscopic mechanisms of single-link addition
control the dynamics of the size of the largest cluster and impact
the type of transition. In particular, the exponent $\beta$ tells
in how far single link additions change macroscopic connectivity.
For generic competitive processes $\beta$ is smaller than for non-competitive
ones (see Fig. \ref{Fig:dcmaxDecay}), but our numerical and analytic
results indicate that they are still distinct from zero. Only processes
with global competition or other extreme forms of competition yield
$\beta=0$ and thus a discontinuous gap $\DCmax$ induced by single
link addition. Others, more generic processes, typically exhibit $\beta>0$
and thus a weakly discontinuous transition. 

It is important to note that percolation processes with only moderate
competition may already yield very small positive exponents and thus
essentially macroscopic gaps (see Fig.~\ref{Fig:originalDynamics}).
Here we used {}``essentially macroscopic'' to mean that (a) the
addition of single links in systems of physically large size induces
gaps that are of relevant size (substantial fraction of system size)
and that (b) the gap sizes increase with stronger competition (e.g.
increasing k) yielding a decreasing exponent $\beta\rightarrow0$
as $k\rightarrow N$. As a consequence, even processes actually exhibiting
weakly discontinuous transitions may display large gaps in systems
of physically relevant size (compare with Fig.~\ref{Fig:dcmaxDecay}).
For instance, if $\beta=0.02$, a system of macroscopic, but finite
size $N=10^{23}$ exhibits a gap of $\DCmax/N\sim N^{-\beta}\approx0.35$
although formally $\DCmax/N\rightarrow0$ as $N\rightarrow\infty$.
For many real processes with already moderate forms of competition,
we expect exponents $\beta$ close to zero, and thus conjecture that
single links may have a strong impact onto how such a network becomes
connected.

In summary, our results demonstrate how in competitive percolation,
keeping the growth rate of the largest cluster small, strengthens
the impact of single link additions that merge smaller clusters. Growing
(i) and overtaking (ii) markedly distinguish the microscopic dynamics
in systems exhibiting competitive percolation. The more largest cluster
growth is suppressed, the more relevant the discontinuous gap becomes
in large systems of given finite size.    Single link additions
may then induce an essentially macroscopic gap even for weakly discontinuous
transitions if competition is sufficiently strong.

Interestingly, a protein homology network has recently been identified
\cite{RozenfeldProteinHomologyExplosive09} that displays macroscopic
features akin to explosive percolation. Individual links may also
induce abrupt changes in several other growing networked systems,
possibly with severe consequences for the systems' dynamics and function
(compare to \cite{Spencer2008,Spencer2010,TimmeEPL2006SCCs}). For
instance, growing one or a few additional synaptic connections in
a neuronal circuit may strongly alter the global connectivity and
thus the overall activity of the circuit \cite{BreskinPRL06,SorianoPNAS08};\textbf{
}specific infected individuals traveling to one but not another location
may drastically change the patterns of infectious diseases \cite{HufnagelPNAS04};
and the macroscopic properties of complex systems exhibiting competitive
aggregation dynamics of physical or biological units may exhibit abrupt
phase transitions induced by a small set of specific individual bonds
newly established, compare, e.g.~\cite{CruzPNAS,DinsmorePRL2006StructureColloids}.
Our study thus does not only provide recipes (by looking for certain
competitive cluster formation) to identify real systems that could
exhibit a (weakly) discontinuous pericolation transition, but also
shows that and how single link additions in such systems may induce
discontinuous gaps, and in turn a collective, very abrupt change of
structure and dynamics. 

The current study answers how single-link dynamics underlies competitive
percolation in general, but does not tell how single link additions
are actually generated and controlled in any given real system. Future
work must bridge this gap and establish how competitive percolation,
and in particular the creation of essentially macroscopic jumps due
to single link additions, is influenced by predefined structure, e.g.
for percolation processes on lattices and in geometrical or topological
confinement occurring in nature \cite{Cho2009,RadicciExplosiveScaleFreePRL09,ZiffExplosive2DLatticePRL09,RozenfeldProteinHomologyExplosive09,Friedman2009}.\textbf{ }

\textbf{Acknowledgments: }We thank Nigel Goldenfeld and Ido Kanter
for fruitful discussions. M.T. acknowledges support by the Federal
Ministry of Education and Research (BMBF) Germany, under Grant No.
01GQ0430 and by the Max Planck Society.

\textbf{Author contributions:} All authors conceived and designed
the research, contributed analysis tools, and analyzed the data. J.N.
performed the numerical experiments. All authors worked out the theory
and wrote the manuscript.

\textbf{Supplementary information} accompanies this article on www.nature.com/naturephysics.
Reprints and permissions information is available online at http://npg.nature.com/reprintsandpermissions.
Correspondence and requests for materials should be addressed to M.T.

\textbf{Competing interests statement:} The authors have declared
that no competing interests exist. 

\bibliographystyle{naturemag}
\bibliography{percolation_bib_v2,competitionv41}

\end{document}